\documentstyle[aps]{revtex}

\begin{document}
\title{Macroscopic quantum coherence in mesoscopic ferromagnetic systems}
\author{S. P. Kou$^{1,2}$, J. Q. Liang$^{1,2}$ , Y.B. Zhang$^{1.2}$ and F. C. Pu$%
^{1,3}$}
\address{$^1$Institute of Physics and Center for Condensed Matter Physics, Chinese\\
Academy of Sciences, Beijing 100080, P. R. China\\
$^2$Department of Physics, Shanxi University, Shanxi 030006, P. R., China\\
$^3$Department of Physics, Guangzhou Normal College, Guangzhou 510400, P.}
\maketitle

\begin{abstract}
In this paper we study the Macroscopic Quantum Oscillation ( MQO) effect in
ferromagnetic single domain magnets with a magnetic field applied along the
hard anistropy axis. The level splitting for the ground state, derived with
the conventional instanton method, oscillates with the external field and is
quenched at some field values. A formula for quantum tunneling at excited
levels is also obtained. The existence of topological phase accounts for
this kind of oscillation and the corresponding thermodynamical quantities
exhibit similar interference effects which resembles to some extent the
electron quantum phase interference induced by gauge potential in the
Aharonov-Bohm effect and the $\Theta $-vacuum in Yang-Mills field theory..

PACS number(s): 75.10.Jm, 03.65.Sq, 73.40.Gk, 75.30.Gw.
\end{abstract}

\section{Introduction}

The remarkable macroscopic quantum phenomena ( MQP ) of nano-magnetic
particles at sufficiently low temperatures stimulated considerable research
activities in recent years\cite{1,2,3}. In single-domain particles, MQP
represents quantum transitions of the magnetic moment {\bf M} between easy
directions in a single-domain ferromagnetic (FM) grain and of the N\'eel
vector {\bf L} in a quantum superposition of AFM sublattices. It has been
pointed out that there is a crossover at $T_B$, above which the transition
process is dominated by thermal hoping and the transition rate follows
Arrhenius law $\Gamma =\nu _0\exp [-U/k_BT]$ where $U$ is the energy barrier
and $\nu _0$ is an attempt frequency, below $T_B$ the transitions occur by
quantum tunneling and a temperature independent rate $\Gamma =\nu _0\exp
[-B] $ is expected. The crossover temperature is defined as $T_B=\frac U{Bk_B%
}$. Macroscopic quantum coherence (MQC) phenomena have been studied
extensively for its exotic characters far from that of classical systems. In
1992, Awschalom et al.\cite{4} observed evidences of resonant tunneling in
dense ferritin solutions in which each particle contains about 4500 $Fe^{3+}$
spin 5/2 ions below 200mK. Recently remarkable quantization phenomena were
observed in $Mn_{12}Ac$\cite{5} that are also understood as a kind of
resonant magnetization tunneling$.$ And resonant tunneling evidences are
also shown in $Fe_8$ as well as $CrR_6$ where $R$ stand for $Mn$ or $Ni$\cite
{6,7}.

It has been shown that for a wide range of magnetic systems, quantum
tunneling is completely suppressed if the total spin is a half integer but
is allowed for integer-spin particles which is called the ''spin-parity
effect'' \cite{8,9,10} and related to Kramers' degeneracy.. The
''spin-parity effect'' is a direct result of the topological phase $e^{i\dot{%
\phi}S}$ \cite{11} and as such does not occur in previous Lagrangian
formulations of ''micromagnetics'', which study the statics and
thermodynamics of continuous magnetization fields. In this paper we study a
kind of MQC phenomena resulting from the topological phase, which we call
macroscopic quantum oscillation(MQO). This kind of oscillation has been
proposed in both ferromagnetic (FM) grains and antiferromagnetic molecular
magnets in the presence of magnetic fields\cite{12,13,14}. It has been
pointed that in ferromagnetic particles with biaxial symmetry and an
external field along the hard axis the tunneling rate-quenching need not be
related to Kramers' degeneracy\cite{12}. Using spin-coherent-state path
integrals, Garg first showed that the tunneling splitting is quenched\cite
{12} whenever $H=\frac{2K_1\sqrt{1-\frac{K_2}{K_1}}}{g\mu _B}\left(
S-n-1/2\right) $ where $K_1$, $K_2$ are the anisotropy constants with $%
K_1>K_2$, $S$ is the total spins of the particles and $n$ is an integer. In
this paper we solve the paths from the classical action and derive a formula
for the level splitting as a function of field, from which his specific
field values can be read off directly. Following ref. (\cite{14}), we
calculate the tunneling dynamics of the magnetic moment of FM particles
which may be measured via the static magnetization and the specific heat.
Here we shall study a spin model with biaxial anisotropy in detail and point
out that quantum oscillations are of the Aharonov-Bohm-type, similar to that
of the $\theta $ vacuum in a Yang-Mills field but different from that in $%
Mn_{12}Ac.$

In Section 2 we show that there is Aharonov-Bohm-type MQO in spin systems
with a uni-axial anisotropy. In section 3 we discuss a spin model with
bi-axial anisotropy and find that the energy level splitting of the ground
state varies with a vacuum angle $\Theta $, which is a good realization of
the $\Theta $ vacuum in condensed matter. The instanton solution is obtained
through the standard procedure and the energy splitting agrees excellently
with numerical results. We derive a compact formula for the level-splitting
induced by tunneling which is valid for the region of low energy levels. The
results of tunneling effects association with the excited states based on
the LSZ (Lehmann, Symanzik and Zimmermann) procedure of field theory are
restricted in the low energy region \cite{15,16,17,18}. Finally we study the
thermodynamic properties of the tunneling effect in Section 5.

\section{MQO in magnetic grains with a easy-axis anisotropy}

We consider the following Hamiltonian operator of a ferromagnetic particle
with uniaxis anisotropy\cite{20} 
\begin{equation}
{\cal H}=K\hat{s}_z^2-g\mu _BH\hat{s}_z  \label{2.1}
\end{equation}
which describes uniaxis anisotropy and an magnetic field along hard axis
with the anisotropy constants $K>0$. The spin operators $\widehat{s}_i$, $%
i=x,y,z,$ obey the usual commutation relation $\left[ \widehat{s}_i,\widehat{%
s}_j\right] =i\epsilon _{ijk}\widehat{s}_k$ (using natural units
throughout). The Hamiltonian is exactly diagonal in terms of the eigenstates 
$\hat{s}_z$ and we have $E_m=Km^2-g\mu _BHm.$ Instead of using the result,
we express the partition function as a spin-coherent-state path integral for
large spins $S\gg 1$\cite{21} 
\begin{equation}
Z=T_re^{-\beta {\cal H}}=\int {\cal D}\{\mu (\stackrel{\rightarrow }{n}%
)\}e^{-S_E}  \label{2.2}
\end{equation}
where ${\cal D}\{\mu (\stackrel{\rightarrow }{n})\}=\prod_{k=1}^{M-1}\frac{%
\sin \theta _kd\phi _kd\theta _k}{2\pi }$ and 
\begin{equation}
S_E=\int_0^{\beta \hbar }d\tau [iS\dot{\phi}(1-\cos \theta )+KS\left(
S+1\right) \cos ^2\theta -g\mu _BHS\cos \theta ].  \label{2.3}
\end{equation}
Although the method from spin-coherent-state seems redundant, our purpose in
this part are to compare this simple model with the model in next part.
After integrating over $\cos \theta ,$ we map the spin system onto a
particle problem with Lagrangian 
\begin{equation}
{\cal L}=\frac{m\dot{\phi}^2}2+i\Theta \dot{\phi}  \label{2.4}
\end{equation}
where $m=\frac 1{2K}$ and $\Theta =S\left( 1-\frac{g\mu _BH}{2KS}\right) $.
This Lagrangian is just similar to that of eq.(3.1) in ref.(\cite{22}). The
last term of ${\cal L}$ is the total imaginary time derivative and has no
effect on the classical equation of motion, though it alters the canonical
momentum $\Pi _\phi =m\stackrel{.}{\phi }+i\Theta .$ This magnetic model
with uniaxial anisotropy in an external field resembles an electron moving
on a conducting ring that is crossed by a flux $\Phi $ which shows
remarkable Aharonov-Bohm effect \cite{23}. With imaginary time $\tau =it$
variable, the periodic instanton configurations are classical solutions
which minimize the Euclidean action under the boundary condition $\phi
_n(\tau +\beta )=\phi _n(\tau )+2\pi n$ 
\begin{eqnarray}
\phi _n &=&\frac{2\pi n}\beta \tau ,  \label{2.5} \\
S_E &=&s_0n^2+i2n\pi \Theta  \nonumber
\end{eqnarray}
where $n$ is the winding number characterizing homotopically nonequivalent
class and $s_0=\pi ^2k_BT/K.$ The Euclidean functional integral of the
partition function contains an additional summation over the homotopic
number 
\begin{equation}
Z=\sum\limits_{n=-\infty }^\infty Z_n=\Theta _3(\pi \Theta ,\exp (-s_0))
\label{2.6}
\end{equation}
where $\Theta _3(v,q)$ is the Jacobi theta function oscillating with $\Theta 
$. In the following we shall be interested only in that part of the ground
state energy which depends on the $\Theta $ 
\begin{eqnarray}
E_0 &=&-k_BT\ln Z\text{ as }T\rightarrow 0  \label{2.7} \\
\ &=&-\frac{\left( g\mu _BH\right) ^2}{4K}+\frac 1{2m}\{\{\Theta \}\}^2 
\nonumber
\end{eqnarray}
where $\{\{x\}\}$ is the fractional part of x to the nearest integer. This
simple model shows remarkable similarity to $\theta $ vacuum of the
non-Abelian gauge field - both of them are a manifestation of vacuum-vacuum
tunneling and angle $\Theta $ can be made as an external variable parameter
in stead of a dynamic one.

>From the partition function we find the oscillation of the magnetic momentum 
$\frac{\partial M}{\partial H}$ with external field $H_z$ along the hard
axis 
\begin{eqnarray}
\frac{\partial M}{\partial H} &=&{\bf -}\frac{k_BT\partial ^2\ln Z}{\partial
H^2}  \label{2.8} \\
\ &\rightarrow &\frac{\left( g\mu _B\right) ^2}{2K}+g\mu _B\left[ \frac 12+%
\frac 12\left( -1\right) ^{\left\{ \Theta \right\} }\right] \text{ as }%
k_BTm\ll 1  \nonumber \\
\ &\rightarrow &\frac{\left( g\mu _B\right) ^2}{2K}-2k_BT\left( \frac{\pi
g\mu _B}K\right) ^2\cos \left( 2\pi \Theta \right) \exp (-s_0/\hbar )\text{
as }k_BTm\gg 1.  \nonumber
\end{eqnarray}
where $\left\{ x\right\} $ is the integer part of x to the nearest integer.
The result shows the regular intervals of magnetic field $\triangle H=\frac{%
2K}{g\mu _B}$. Fig.1 shows the magnetization curves at zero temperature $M-H$
: curve I is magnetization curve for magnetic particles with integer spin,
curve II with half integer spin and curve III is the classical one which has
no oscillations. This kind of quantum behaviors is essentially a kind of
''Aharonov-Bohm effect'' in mesoscopic magnetic systems resulting from Berry
phase. Recently quantum steps in the hysteresis loop of $M_{n12}Ac$ crystal
have been observed at regular intervals of magnetic field with period $%
\triangle H=\frac K{g\mu _B}$. However the Hamiltonian of $M_{n12}Ac$ is $%
H=-K\widehat{s}_z^2-g\mu _BH\widehat{s}_z+H^{\prime }$ where $H^{\prime }$
is the perturbation term and can't communicate with other terms, such as $%
\widehat{s}_{\pm }^4$ and spin-phonon interaction. Even without the term $%
H^{\prime }$, this model can't be mapped onto the above model due to the
opposite sign of mass, $m=\frac{-1}{2K}$. The difference between the two
kinds of MQPs is obvious : in the model here MQP is essential as a kind of
''Aharonov-Bohm effect'' in mesoscopic magnetic systems from Berry phase
with a period $\triangle H=\frac{2K}{g\mu _B}$; while in $M_{n12}Ac$ MQC is
a effect of energy level-crossing and the period $\triangle H=\frac K{g\mu _B%
}$ ( The perturbation term $H^{\prime }$ is the key to understand the exotic
phenomenon in $M_{n12}Ac$)$.$

\section{MQO in magnetic grains with a biaxial anisotropy}

Let us consider the following Hamiltonian operator of a ferromagnetic
particle with XOY easy plane anisotropy and $x$ easy axis with a magnetic
field applied along $z$ axis\cite{12} 
\begin{equation}
\widehat{H}=K_1\widehat{s}_z^2+K_2\widehat{s}_y^2-g\mu _BH\widehat{s}_z
\label{3.1}
\end{equation}
where $K_1>K_2>0.$ Starting from the coherent state representation of the
time evolution operator with Hamiltonian 
\begin{equation}
E(\theta ,\phi )=K_1\cos ^2\theta +K_2\sin ^2\theta \sin ^2\phi -g\mu
_BH\cos \theta  \label{3.2}
\end{equation}
and integrating cos$\theta ,$ we map the spin system onto a particle problem
with the effective Lagrangian 
\begin{equation}
{\cal L}_{eff}=\frac 12m(\phi )\stackrel{.}{\phi }^2-V(\phi )+\Theta (\phi )%
\stackrel{.}{\phi }  \label{3.3}
\end{equation}
where the position dependent mass $m(\phi )$ and angle $\Theta (\phi )$ are
defined by 
\begin{eqnarray}
m(\phi ) &=&\frac 1{2K_1(1-\lambda \sin ^2\phi )},  \label{3.4} \\
\text{ }\Theta (\phi ) &=&S(1-\frac H{2K_1S(1-\lambda \sin ^2\phi )}), 
\nonumber \\
V(\phi ) &=&K_2S\left( S+1\right) \sin ^2\phi -\frac{\left( g\mu _BH\right)
^2\lambda \sin ^2\phi }{4K_1(1-\lambda \sin ^2\phi )}  \nonumber
\end{eqnarray}
respectively. Here $S$ is the total spin of the particles and $\lambda =%
\frac{K_2}{K_1}$. The term $\Theta (\phi )\stackrel{.}{\phi }$ has no effect
on the classical motion equation and comes from the Berry phase.

\subsection{Level\ Splitting of ground states}

Now we derive the level splitting of a giant spin particle in the large spin
limit. Instantons in field theory of $0+1$ dimensions viewed as
pseudoparticles with trajectories existing in barriers, are therefore
responsible for tunneling. Since instanton trajectories interpolate between
degenerate vacua and satisfy vacuum boundary conditions, the instanton
method is suitable only for the calculation of tunneling parameter $\Delta
\epsilon _0$ between neighboring vacua. In the following we first consider
tunneling at the vacuum level (i. e. $m=0$), which leads to the level
splitting of the ground state energy, i.e. $2\Delta \epsilon _0$. Passing to
imaginary time by Wick rotation $\tau =it,\beta =iT,$ the amplitude for
tunneling from the initial well, say that with $n=0$ (and $\phi _i=0$), to
the neighboring well with $n=1$ (and $\phi _f=\pi $), is obtained by 
\begin{equation}
<0,\Phi _n=\pi \mid e^{-\beta {\cal \hat{H}}}\mid 0,\Phi _{n+1}=0>={\cal K}%
_E(\phi _f=\pi ,\beta ;\phi _i=0,0)  \label{3.5}
\end{equation}
where $\mid 0,\Phi _n>$ is the perturbation wave function of the ground
state at the well denoted by $\Phi _n$. Since the classical solution $\phi
_c $ satisfies the equation of motion $\frac 12m(\phi _c)\stackrel{.}{\phi }%
^2-V(\phi _c)=0,$ we have 
\begin{equation}
\phi _c=\arccos [\sqrt{\frac{(1-\lambda -u)\tanh ^2\omega _0\tau }{%
1-u-\lambda \tanh ^2\omega _0\tau }}]  \label{3.6}
\end{equation}
where $\omega _0=\sqrt{4K_1K_2S\left( S+1\right) (1-u)}$ and $u=\frac{(g\mu
_BH)^2}{4K_1^2S\left( S+1\right) }$ with non-vanishing topological charge
and the Euclidean action is 
\begin{eqnarray}
S_c &=&\sqrt{S\left( S+1\right) }\ln \left[ \frac{\sqrt{(1-u)}+\sqrt{\lambda 
}}{\sqrt{(1-u)}-\sqrt{\lambda }}\right]  \label{3.7} \\
&&\ \ \ \ \ \ \ -\sqrt{S\left( S+1\right) }\sqrt{\frac u{1-\lambda }}\ln
\left[ \frac{1+\sqrt{\xi }}{1-\sqrt{\xi }}\right]  \nonumber
\end{eqnarray}
where $\xi =\frac{\lambda u}{(1-\lambda )(1-u)}$ and $u<(1-\lambda )$. When $%
H\rightarrow 0$, we have 
\begin{equation}
\cos ^2\phi _c=\frac{1-\tanh ^2\omega _0\tau }{1-\lambda \tanh ^2\omega
_0\tau }  \label{3.8}
\end{equation}
which is exactly the vacuum instanton solution derived previously in the
absence of magnetic field\cite{2,17,18}.

The prefactor can be evaluated with the stationary phase method by expanding 
$\phi $ around the instanton trajectory $\phi _c$ such that $\phi =\phi
_c+\eta $, where $\eta $ is the small fluctuation with boundary conditions $%
\eta (\beta )=\eta (0)=0$. Up to the one-loop approximation we have 
\begin{equation}
{\cal K}_E=e^{-S_c}I  \label{3.9}
\end{equation}
where 
\begin{equation}
I=\int_{\eta (0)=0}^{\eta (\beta )=0}{\cal D}\eta e^{-\delta S_E}
\label{3.10}
\end{equation}
is the fluctuation functional integral with the fluctuation action 
\begin{equation}
\delta S_E=\int_0^\beta \eta \hat{M}\eta d\tau  \label{3.11}
\end{equation}
where 
\begin{equation}
\hat{M}=-\frac 12\frac d{d\tau }m(\phi _c)\frac d{d\tau }+\tilde{V}(\phi _c)
\label{3.12}
\end{equation}
with 
\begin{equation}
\tilde{V}(\phi _c)=\frac 12[-m^{\prime }(\phi _c)\ddot{\phi}_c-\frac 12%
m^{\prime \prime }(\phi _c)\dot{\phi}_c^2+V^{\prime \prime }(\phi _c)]
\label{3.13}
\end{equation}
Here $\hat{M}(\phi _c)$ is the operator of the second variation of the
action and $m^{\prime }(\phi _c)={\frac{\partial m(\phi )}{\partial \phi }}%
|_{\phi =\phi _c}$. As in the usual method of evaluating the fluctuation
integral $I$, we expand the fluctuation variable $\eta $ in terms of the
eigenmodes of $\hat{M}$ and set $\eta =\Sigma _nC_n\psi _n$, where $\psi _n$
denotes the $n$-th eigenfunction of $\hat{M}$, and express the result of the
integration as an inverse square root of the determinant of $\hat{M}$. In
view of the time translation symmetry of the equation of motion, the
functional integral ${\cal K}_E$ is not well defined when expanded about the
classical solutions. The translational symmetry results in zero eigenmodes
of the second variation operator $\hat{M}$ of the action. This problem can
be cured by the Faddeev--Popov procedure \cite{24}. Following the procedure
of refs.\cite{26}, the one instanton contribution to the propagator in the
one-loop approximation is calculated as 
\begin{equation}
{\cal K}_E^{(1)}=\beta \frac 4\pi (1-u-\lambda )^{-\frac 12}(1-u)^{\frac 32%
}S^2K_2e^{-\omega _0\beta /2}e^{-S_c}e^{-i\pi \Theta }  \label{3.14}
\end{equation}

To obtain the desired result proportional to $\sinh (\beta \bigtriangleup
E_0/2)$ in eq.(\ref{3.5}), the contributions of the infinite number of
instanton and anti-instanton pairs to the one instanton contribution have to
be taken into account. Interactions among instantons and anti-instantons are
neglected in the dilute instanton--gas approximation. Summing over all
contributions of instantons and anti-instantons, the final result of the
propagator is found to be 
\begin{eqnarray}
{\cal K}_E &=&\lambda ^{\frac 14}(\frac S\pi )^{\frac 12}\left( 1-u\right) ^{%
\frac 14}e^{-\beta \omega _0/2}  \label{3.15} \\
&&\ \ \times \sum\limits_{n,m\geq 0}^{m+n\text{ odd}}\frac 1{m!n!}\left(
2D\right) ^{m+n}e^{-S_c\left( m+n\right) }e^{-i\pi \Theta \left( m-n\right) }
\nonumber \\
\ &=&\lambda ^{\frac 14}(\frac S\pi )^{\frac 12}\left( 1-u\right) ^{\frac 14%
}e^{-\beta \omega _0/2}  \nonumber \\
&&\ \ \ \ \ \sinh [\beta D\cos \left( \pi \Theta \right) e^{-S_c}]  \nonumber
\end{eqnarray}
where $D=2^3\{\frac{K_1K_2}{(1-u-\lambda )\pi }\}^{\frac 12}(1-u)^{\frac 54%
}\lambda ^{\frac 14}S^{\frac 32}$ and $\Theta =S\left( 1-\frac{g\mu _BH}{%
2SK_1\left( 1-\lambda \right) ^{1/2}}\right) .$ Compared with eq.(\ref{3.5})
the level shift is seen to be $2\triangle \epsilon _0\cos \left( \pi \Theta
\right) $ (note that this is the shift of a single level), the tunneling
splitting is 
\begin{equation}
\triangle E_0=2\triangle \epsilon _0\cos \left( \pi \Theta \right)  \nonumber
\end{equation}
where 
\begin{equation}
\triangle \epsilon _0=8\{\frac{K_1K_2}{(1-u-\lambda )\pi }\}^{\frac 12%
}(1-u)^{\frac 54}\lambda ^{\frac 14}S^{\frac 32}e^{-S_c/\hbar }.
\label{3.16}
\end{equation}
The tunneling splitting is quenched $\left( \triangle E_0=0\right) $
whenever $\Theta =k+1/2$ or $H=\frac{2\sqrt{1-\lambda }K_1}{g\mu _B}%
(S-k-1/2)/S$ where $k$ is an integer and this is the result of topological
phase : there is destructive interference between any path $\phi (\tau )$
and $\theta (\tau )$ with its time reversal pair $-\phi (\tau )$ and $\pi
-\theta (\tau ).$ The result is in good agreement with earlier works by Garg%
\cite{12}. The original Hamiltonian operator in a magnetic field along $z$
axis has no time-reversal symmetry for $g\mu _BH\widehat{s}_z\rightarrow
-g\mu _BH\widehat{s}_z$ as $t\rightarrow -t$ where $t$ is time. For this
reason, this quenching in magnetic field has nothing to do with Kramers'
degeneracy and is different from the quenching with half-integer spins in
the absence of magnetic field.

\subsection{Level splitting of low-lying levels and the LSZ method}

A formula suitable for a quantitative of the tunneling effect at excited
levels is obtained with the Lehmann, Symanzik and Zimmermann (LSZ) method in
field theory in which the tunneling is viewed as the transition of $n$
bosons induced by the usual(vacuum) instanton. The idea of a tunneling
transition from one side of a potential barrier to the other has recently
also been linked with the LSZ reduction mechanism of a transition from the
asymptotic in-state to asymptotic out-states. In the following we use the
LSZ reduction procedure in a modified way in order to calculate the
tunneling in the one-instanton sector for the effective potential of the
spin system, including the contribution of quantum fluctuations up to the
one-loop approximation.

We recall first the case of a one-dimensional harmonic oscillator described
by the Hamiltonian 
\begin{equation}
H=\frac 12p^2+\frac 12\omega _0^2q^2  \label{3.18}
\end{equation}
for mass $m=1$. Here $q$ and $p$ are dynamical observable which become
operators when subjected to the Heisenberg algebra of ordinary canonical
quantization. The solution of the Heisenberg equation of motion, $\ddot{q}%
+\omega _0^2q=0$, then becomes 
\begin{equation}
q(t)=\frac 1{\sqrt{2\omega _0}}[ae^{-i\omega _0t}+a^{\dagger }e^{i\omega
_0t}]  \label{3.19}
\end{equation}
where $a,a^{\dagger }$ are time-independent operators defined by the initial
($t=0$) values of $q$ and $p$, i.e. 
\begin{eqnarray}
q(0) &=&\frac 1{\sqrt{2\omega _0}}[a+a^{\dagger }],  \label{3.20} \\
p(0) &=&\frac{-i\omega _0}{\sqrt{2\omega _0}}[a-a^{\dagger }]  \nonumber
\end{eqnarray}
The operators $a,a^{\dagger }$ can be obtained from $q(t),p(t)=\dot{q}(t)$.
Thus 
\begin{eqnarray}
a^{\dagger } &=&-\frac i{\sqrt{2\omega }}e^{-i\omega _0t}[\dot{q}(t)+i\omega
_0q(t)]  \label{3.21} \\
\ &\equiv &-\frac i{\sqrt{2\omega }}e^{-i\omega _0t}\stackrel{%
\leftrightarrow }{\frac \partial {\partial t}}q(t)  \nonumber
\end{eqnarray}
and $a$ follows with complex conjugation. One should note the extra $-$ in
the definition of the symbol $\stackrel{\leftrightarrow }{\frac \partial {%
\partial t}}$ when acting to the left. Operators of this type are well-known
in the literature. Because the ''harmonic oscillator'' approximation is only
useful of low lying levels, the results from LSZ are restricted to low lying
levels.

We now consider the $(0+1)$ dimensional theory defined by the Euclidean
Lagrangian eq.(\ref{3.4}). Crucial aspects of the LSZ procedure are its
asymptotic conditions which require the theory to have an interpretation in
terms of observables for stationary in coming and outgoing states. We can
simulate such a situation here artificially by imagining the central barrier
of the potential to be extremely high and the neighboring wells $"-"$ and $%
"+"$ on either side to be extremely far apart. We therefore construct
appropriate functions $\phi _{\pm }(\tau )$ which become 
\begin{equation}
\phi _{+}=\pi -\phi _c,\qquad \phi _{-}=\phi _c  \label{3.22}
\end{equation}
such that the interaction fields vanish in their respective asymptotic
regions, i.e. 
\begin{equation}
\lim_{\tau \rightarrow +\infty }\phi _{+}(\tau )=0,\qquad \lim_{\tau
\rightarrow -\infty }\phi _{-}(\tau )=0  \label{3.23}
\end{equation}
The subscripts $"-"$ and $"+"$ here denote the wells with minima at $\Phi
_0=0$ and $\Phi _1=\pi $ respectively. The Euclidean creation and
annihilation operators $\widehat{a}_{\pm }^{+}$ and $\widehat{a}_{\pm }$
which create and annihilate an effective boson in wells $"+"$ and $"-"$
respectively are related to the interaction field operators $\phi _{\pm }$
by 
\begin{eqnarray}
\widehat{a}_{\pm }^{\dagger }(\tau ) &:&=\sqrt{\frac{2m_0}{\omega _0}}%
e^{-\omega _0\tau }\stackrel{\leftrightarrow }{\frac \partial {\partial \tau 
}}\phi _{\pm }(\tau ),  \label{3.24} \\[0.0254cm]
\widehat{a}_{\pm }(\tau ) &:&=-\sqrt{\frac{2m_0}{\omega _0}}e^{\omega _0\tau
}\stackrel{\leftrightarrow }{\frac \partial {\partial \tau }}\phi _{\pm
}(\tau )\   \nonumber
\end{eqnarray}
where $:=$ represents definition and $m_0=1/2K_1.$ From the viewpoint of the
LSZ method the transition amplitude between $m$-th degenerate eigenstates in
any two neighboring wells (here for $n=0,1$) is viewed as the transition of $%
m$ bosons induced by the instanton of eq.(\ref{3.6}) and is related to the
tunneling parameter $\Delta \epsilon _m$ by 
\begin{equation}
A_{f,i}^m=<m,\Phi _1|e^{-\beta \widehat{H}}|m,\Phi _0>=e^{-\beta \epsilon
_m}\sinh \beta \Delta \epsilon _m  \label{3.25}
\end{equation}
where $\mid m,\Phi _n>$ is the perturbation wave function of the excited
level at the well denoted by $\Phi _n$ The transition amplitude as well as
the $S$-matrix can be related to the Green's function through the procedure
known as the LSZ reduction technique. To this end we rewrite the transition
amplitude as 
\begin{equation}
A_{f,i}^m=S_{f,i}^me^{-\beta \omega _0}  \label{3.26}
\end{equation}
with $S$-matrix element 
\begin{equation}
S_{f,i}^m=\lim_{%
{\tau ^i\to -\infty  \atop \tau ^f\to \infty }%
}\frac 1{m!}\langle 0|\hat{a}_{+}(\tau _m^f)\ldots \hat{a}_{+}(\tau _1^f)%
\hat{a}_{-}^{\dagger }(\tau _1^i)\ldots \hat{a}_{-}^{\dagger }(\tau
_m^i)|0\rangle  \label{3.27}
\end{equation}
The $S$-matrix element can be evaluated in terms of the Green's function $G$
which arises in its evaluation. Thus 
\begin{eqnarray}
S_{f,i}^m &=&\lim_{%
{\tau ^i\to -\infty  \atop \tau ^f\to \infty }%
}\frac 1{m!}\prod_{l=1}^m\left( -\sqrt{\frac{2m_0}{\omega _0}}e^{\omega
_0\tau _l^f}\stackrel{\leftrightarrow }{\frac \partial {\partial \tau _l^f}}%
\right) \left( \sqrt{\frac{2m_0}{\omega _0}}e^{-\omega _0\tau _l^i}\stackrel{%
\leftrightarrow }{\frac \partial {\partial \tau _l^i}}\right) G  \label{3.28}
\\
\ &=&\frac 1{m!}\prod_{l=1}^m\left( \frac{-2m_0}{\omega _0}\right) e^{\omega
_0(\tau _l^f-\tau _l^i)}\left[ \frac{\partial ^2G}{\partial \tau
_l^f\partial \tau _l^i}+\omega _0\left( \frac{\partial G}{\partial \tau _l^f}%
-\frac{\partial G}{\partial \tau _l^i}-\omega _0G\right) \right]  \nonumber
\end{eqnarray}
where the $2m$-point Green's function is defined as usual, i.e. 
\begin{equation}
G=\langle 0|\hat{\phi}_{+}(\tau _m^f)\ldots \hat{\phi}_{+}(\tau _1^f)\hat{%
\phi}_{-}(\tau _1^i)\ldots \hat{\phi}_{-}(\tau _m^i)|0\rangle  \label{3.29}
\end{equation}
We evaluate $G$ by inserting complete sets of states of final and initial
field configurations $\phi _f,\phi _i.$ Thus 
\begin{equation}
G(\tau ^f,\tau ^i)=\phi _{+}(\tau ^f)\phi _{-}(\tau ^i)A_{f,i}^0
\label{3.30}
\end{equation}
which vanishes in the limit $\tau ^i\rightarrow -\infty ,\tau ^f\rightarrow
\infty ,$ due to eq. (\ref{3.23}). Thus in eq. (\ref{3.28}) the only
nonvanishing contribution in these limits results from the second
derivative, then 
\begin{equation}
\frac{\partial ^2G}{\partial \tau ^f\partial \tau ^i}=\frac{\partial \phi
_{+}(\tau ^f)}{\partial \tau ^f}\frac{\partial \phi _{-}(\tau ^i)}{\partial
\tau ^i}A_{f,i}^0  \label{3.31}
\end{equation}
the $S$-matrix element for the transition of $m$ bosons is thus 
\begin{equation}
S_{f,i}^m=\frac 1{m!}\prod_{l=1}^m\left\{ \left( -\frac{2m_0}{\omega _0}%
\right) \left[ \frac{d\phi _{+}(\tau _l^f)}{d\tau _l^f}\frac{d\phi _{-}(\tau
_l^i)}{d\tau _l^i}\right] \right\} A_{f,i}^0  \label{3.32}
\end{equation}
The transition amplitude between degenerate ground state can be calculated
from the definition, eq. (\ref{3.18}), in terms of the tunneling parameter $%
\Delta \epsilon _0,$i.e. $A_{f,i}^0=\beta \Delta \epsilon _0\cos \left(
\Theta \pi \right) e^{-\beta \omega _0/2}.$ Then we have 
\begin{equation}
A_{f,i}^m=\frac 1{m!}q^me^{-\omega _0m\beta }A_{f,i}^0  \label{3.33}
\end{equation}
where $q=\frac{8s\lambda ^{1/2}(1-u)^{3/2}}{1-\lambda -u}.$ The tunneling
parameter at $m$-th excited state is seen to be 
\begin{equation}
\Delta \epsilon _m=\frac 1{m!}q^m\Delta \epsilon _0\cos \left( \Theta \pi
\right)  \label{3.34}
\end{equation}
where $\triangle \epsilon _m$ is the usual overlap integral or simply the
level shift due to tunneling through any one of the barriers. For $q=\frac{%
8S\lambda ^{1/2}(1-u)^{3/2}}{1-\lambda -u}\gg 1$ the tunneling effect is
much more obvious of higher excited states. The energy levels are 
\begin{equation}
E_{m,\xi }=\epsilon _m-\xi \Delta \epsilon _m\cos \left( \Theta \pi \right)
\label{3.35}
\end{equation}
where $\epsilon _m=\frac{\left( g\mu _BH\right) ^2}{4K_1}+(m+\frac 12)\omega
_0$ and $\xi $ is an integer and here can take only either of the two values
``$0$'' and ``$1$''.

In the context of these investigations the usual terminology of MQC refers
to the resonance between neighboring degenerate wells. There is an essential
difficulty related to the existing theory of quantum tunneling itself in the
absence of an external magnetic field. The argument of the WKB exponential
of the tunneling for a ferromagnetic particle is $2\sqrt{\lambda }s$ with $%
\lambda =\frac{K_2}{K_1}$, $K_1$ and $K_2$ being the hard and medium axis
energies. Recently, evidence for resonant tunneling were observed in $%
Mn_{12}Ac,$ $Fe_8$ and $CrR_6$ ( $R$ equals $Mn$ or $Ni)$\cite{5,6,7}, in
which magnetic clusters have very small spins $\left( S=10-20\right) .$ We
show some data for a ferromagnetic particle in order to demonstrate the
tunneling spliting in magnetic fields $H$ with respect to the result of
tunneling for half integer-spin particles in Fig.2(a) and half integer one
in Fig.2(b). The result of the instanton method is plotted with the solid
line and the numerical result from exact diagonalization is shown by the
dotted line. The level splitting of low-lying levels can be also obtained
with periodic instanton method. For the comparison see the appendices I.

\section{Observation of MQO}

In this part we discuss the possible relevance to experimental test for the
topological phase interference or spin-parity effect in the single domain FM
particles.

The direct way to observe MQOs is to observe the resonant tunneling
frequency. Let's denote $\mid n>_{+}$ and $\mid n>_{-}$ be the eigenstates
of the same energy $E_n$ in the (separate) right and left wells,
respectively. Due to quantum tunneling, the degeneracy is lifted and there
emerges an energy level splitting $\Delta E_n.$ The new eigenstates are $%
\mid n>_o=\frac 1{\sqrt{2}}(\mid n>_{+}-\mid n>_{-})$ with odd parity and $%
\mid n>_e=\frac 1{\sqrt{2}}(\mid n>_{+}+\mid n>_{-})$ with even parity. If
we choose $\mid \Psi (t=0)>_{+}=\mid n>_{+}$as initial state, the
probability of the system being in the other well is $\mid C_{kn}(t)\mid
^2=\mid \ _{-}<n(t)\mid n>_{+}\mid ^2=\sin ^2[\frac{\Delta E_nt}\hbar ].$
The resonant tunneling frequencies $\Gamma =\frac{\Delta E_n}\hbar =\frac 1{%
n!}\mid \cos \left( \pi \Theta \right) \mid \Delta \epsilon _0q^n$ now with
magnetic field.

Since the topological phase interference can be reflected in thermodynamic
quantities, it is reasonable to study the thermodynamic properties ( such as
the specific heat and the magnetic susceptibility ). Then we discuss
thermodynamic behavior of this model at very low temperature $T\sim \Delta
\epsilon _0/k_B$. At such low temperature the system can be regarded as a
typical two-level system 
\begin{equation}
Z=\sum\limits_{m=0,\xi }e^{-\beta E_{m,\xi }}=e^{-\beta \epsilon _0}\left[
e^{-\beta \cos \left( \Theta \pi \right) \Delta \epsilon _0}+e^{\beta \cos
\left( \Theta \pi \right) \Delta \epsilon _0}\right] .  \label{3.36}
\end{equation}
>From the partition function we can calculate the special heat as $%
C_v=k_B\left[ \frac{2\cos \left( \Theta \pi \right) \Delta \epsilon _0}{k_BT}%
\right] ^2\sec $h$^2\left[ \frac{2\cos \left( \Theta \pi \right) \Delta
\epsilon _0}{k_BT}\right] $ which exhibits a characteristic Schottky anomaly%
\cite{14} shown in Fig. 3. The height of the peak is a constant $0.64k_B$
and the correspond temperature $T_{\max }$ exhibits oscillations with a
period $\frac{2K_1\sqrt{1-\lambda }}{g\mu _B}$%
\begin{equation}
T_{\max }=\frac{1.2}{k_B}\mid \cos \left( \Theta \pi \right) \mid \Delta
\epsilon _0.  \label{3.37}
\end{equation}

The magnetization {\bf M\ }and its derivative with respect to magnetic field
are found to be 
\begin{eqnarray}
M &=&\frac{\left( g\mu _B\right) ^2H}{2K_1}+\left[ 2\cos \left( \Theta \pi
\right) \frac{\partial \Delta \epsilon _0}{\partial H}-2\Delta \epsilon
_0\sin \left( \Theta \pi \right) \frac{\partial \theta }{\partial H}\right]
\label{3.38} \\
&&\ \times \tanh \left( \frac{2\cos \left( \Theta \pi \right) \Delta
\epsilon _0}{k_BT}\right)  \nonumber \\
\ &\simeq &\frac{\left( g\mu _B\right) ^2H}{2K_1}-\frac{\sin \left( 2\Theta
\pi \right) \Delta ^2\epsilon _0}{k_BT}\frac{g\mu _B\pi }{2K_1S\sqrt{%
1-\lambda }}  \nonumber
\end{eqnarray}
and 
\begin{equation}
\frac{\partial M}{\partial H}=\frac{\left( g\mu _B\right) ^2}{2K_1}-\frac{%
2\cos \left( 2\Theta \pi \right) \Delta ^2\epsilon _0}{k_BT}\left( \frac{%
g\mu _B\pi }{2K_1S\sqrt{1-\lambda }}\right) ^2  \label{3.39}
\end{equation}
respectively. The period of oscillation in $\frac{\partial M}{\partial H}$
is $\frac{2K_1\sqrt{1-\lambda }}{g\mu _B}$.

When the temperature is higher $\Delta \epsilon _0\ll k_BT<\omega _0$, the
exited energy levels may give contribution to the partition function and we
have the partition function as the following form 
\begin{eqnarray}
Z &=&\sum\limits_{m,\xi }\exp \{-\beta \left[ \frac{\left( g\mu _BH\right) ^2%
}{4K_1}+(m+\frac 12)\omega _0+2\Delta \epsilon _m\cos (\Theta \pi +\xi \pi
)\right] \}  \label{3.40} \\
\ &\simeq &Z_0+2\left[ \beta \Delta \epsilon _0\cos \left( \Theta \pi
\right) \right] ^2\exp \left[ -\frac{\beta \left( g\mu _BH\right) ^2}{4K_1}%
\right] I_0\left[ 2qe^{-\beta \omega /2}\right]  \nonumber
\end{eqnarray}
where $I_0\left[ x\right] =\sum \frac 1{\left( n!\right) ^2}x^{2n}$ is the
modified Bessel function. We define a characteristic temperature $\tilde{T}$
that is solution to the following equation $qe^{-\omega _0/2k_B\tilde{T}%
}\sim 1$. Only below $\tilde{T}\sim \frac{\omega _0}{2k_B\ln \left( q\right) 
}$ can the system regarded as a two-level system.

>From the partition function we obtain the free energy 
\begin{eqnarray}
F &=&-\frac 1\beta \ln Z  \label{3.41} \\
\ &\simeq &-\frac 1\beta \ln Z_0--2\left[ \Delta \epsilon _0\cos \left(
\Theta \pi \right) \right] ^2F\left( T,H\right)  \nonumber
\end{eqnarray}
where $F\left( T,H\right) =\frac{I_0\left[ 2qe^{-\omega _0/2k_BT}\right] }{%
k_BT\left( e^{\omega _0/k_BT}-1\right) }$. The magnetization ${\bf M}${\bf \ 
}and its derivative with respect to magnetic field are found to be 
\begin{equation}
M\simeq \frac{\left( g\mu _B\right) ^2H}{2K_1}-\frac{2\pi g\mu _B\sin \left(
2\Theta \pi \right) \Delta ^2\epsilon _0}{2K_1S\sqrt{1-\lambda }}F\left(
T,H\right) \text{ }  \label{3.42}
\end{equation}
and 
\begin{equation}
\frac{\partial M}{\partial H}=\frac{\left( g\mu _B\right) ^2}{2K_1}-2\cos
\left( 2\Theta \pi \right) \Delta ^2\epsilon _0\left( \frac{g\mu _B\pi }{%
2K_1S\sqrt{1-\lambda }}\right) ^2F\left( T,H\right)  \label{3.43}
\end{equation}
respectively.

Our method is only useful at low temperatures and it is not reliable above
the crossover temperature $T_B$. Roughly speaking the picture of crossover
phenomena of transition processes through the barrier can be outlined as
follows. At very low temperatures (near zero) the transition is dominated by
instantons or bounces which satisfy the vacuum boundary condition and are
responsible for tunneling at ground state. At high temperature the
transition is mainly due to the thermal activation (over barrier transition)
and follows the Arrehenius law($\sim \exp [-\frac{\Delta U}{k_BT}]$) with $%
\Delta U$ being the barrier height. The transition at high temperature is
induced by sphelaron from the viewpoint of pseudoparticle method. The
crossover occurs at certain temperature $T_B$ above which there is only
sphelaron and below which only periodic instantons contribute to the
imaginary part of partition functions. Above the crossover temperature the
effects of the energy levels above the barrier are more important.

\section{Conclusion}

We have discussed a spin model with biaxial anisotropy and find the energy
splitting of ground state varies with an angle $\Theta $ that is a good
realization of the $\theta $ vacuum structure in condensed matter. In this
paper we show a kind of macroscopic quantum oscillation ( MQC) phenomena in
magnetic grains that is also the direct result of the topological phase and
point out that some thermodynamic parameters (such as special heat $C_v$ or $%
dM/dH$) may show oscillations in a magnetic particles with a easy-plane
anisotropy, occurring in a external field along the symmetry-axis with the
period $\triangle H=\frac{K_1\sqrt{1-\frac{K_2}{K_1}}}{g\mu _B}$. While in a
magnetic field along the medium axis, the small particles also exhibit
quantum interference effects but there is no oscillation of tunneling
splitting with the magnetic field. The energy levels splitting are read as
the following form 
\begin{equation}
\triangle E=\triangle E_A+\triangle E_B\cos \left( 2S\pi \right)  \label{4.1}
\end{equation}
where $\triangle E_A$ and $\triangle E_B$ have no oscillations with magnetic
field\cite{27,28,29}.

\begin{center}
{\bf Appendix 1: Level splitting of low-lying levels and periodic instanton
method}
\end{center}

It is by now known that the periodic instanton method has became a powerful
tool for evaluation of quantum tunneling at entire region of energy. The
classical solution $\phi _c$ satisfies the equation 
\begin{equation}
\frac 12m(\phi _c)\stackrel{.}{\phi }_c^2-V(\phi _c)=-E_{cl}  \label{A2.1}
\end{equation}
which is obtained from by integration once and the integration constant $%
E_{cl}>0$ may be viewed as the classical energy of the pseudoparticle
configuration 
\begin{equation}
\phi _c=\arcsin \sqrt{\frac{1-k^2%
\mathop{\rm sn}%
^2\omega \tau }{1-\lambda k^2%
\mathop{\rm sn}%
^2\omega \tau }}  \label{A2.2}
\end{equation}
where the parameters are 
\begin{equation}
\omega =\omega _0\sqrt{1-n_1^{-2}n_2^2},  \label{11}
\end{equation}
\begin{equation}
k=\sqrt{\frac{n_1^2-1}{n_1^2-n_2^2}}  \label{A2.3}
\end{equation}
and 
\begin{eqnarray}
n_1^2 &=&1/[\frac a{2\lambda }-\sqrt{\frac b\lambda +\frac{a^2}{4\lambda }]},
\label{A2.4} \\
n_2^2 &=&\lambda /[\frac a{2\lambda }+\sqrt{\frac b\lambda +\frac{a^2}{%
4\lambda }}]  \nonumber
\end{eqnarray}
$a=-(u-\frac{4K_1\lambda E}{\omega _0^2}-1)$ and $b=\frac{4K_1E}{\omega _0^2}
$ satisfy the condition $n_1^2>1>n_2^2>0$.

The amplitude for the transition from the left-hand well to the right-hand
well at the energy $E_{cl}$ due to instanton tunnelling can be written as 
\begin{equation}
A_{f,i}^{E_{cl}}=\left\langle E_{cl},\Phi _{n+1}\left| e^{-2HT}\right|
E_{cl},\Phi _n\right\rangle =e^{-2E_{cl}T}\sinh \left( 2T\Delta E_{cl}\right)
\label{A2.5}
\end{equation}
and can also be evaluated with the help of the path-integral method as 
\begin{equation}
A_{f,i}^{E_{cl}}=\int \psi _{E_{cl},n+1}^{*}(\phi _f)\psi _{E_{cl},n}(\phi
_i){\cal K}_E(\phi _f,\tau _f;\phi _i,\tau _i)d\phi _fd\phi _i  \label{A2.6}
\end{equation}
where $\psi _{E_{cl},n}(\phi _i),$ $\psi _{E_{cl},n+1}(\phi _f)$ are wave
functions in the $n$-th, $(n+1)$-th well respectively which extend into the
domain of the barrier. $\Delta E_{cl}$ denotes the energy shift due to
tunneling and is the key quantity to be calculated. Following the procedure
of ref. \cite{15,16,17,18}, up to the one-loop approximation we have 
\begin{equation}
{\cal K}_E=e^{-S_c}I  \label{A2.7}
\end{equation}
and the fluctuation integral can be calculated as 
\begin{equation}
I=\frac 1{\sqrt{2\pi }}\left[ N(\tau _i)N(\tau _f)\int_{-T}^T\frac{d\tau }{%
N^2(\tau )m\left( \phi _c(\tau )\right) }\right] ^{-\frac 12}  \label{A2.8}
\end{equation}
where $N(\tau )\equiv \frac{d\phi _c}{d\tau }$ is the zero-mode of the
periodic instanton solution and the Euclidean action evaluated for the
periodic instanton trajectory is given by 
\begin{eqnarray}
S_c\left( \phi (T),\phi (-T),T\right) &=&\int_{-T}^T\left( \frac 12m(\phi _c)%
\stackrel{.}{\phi }_c^2+V\right) d\tau  \nonumber \\
&=&W\left( \phi (T),\phi (-T),T\right) +2E_{cl}T+i\theta  \label{A2.9}
\end{eqnarray}
where 
\begin{equation}
W\left( \phi (T),\phi (-T),E_{cl}\right) =\frac \omega {\lambda K_1}\left[
K(k)-(1-\lambda k^2)\Pi (\lambda k^2,k)\right]  \label{A2.10}
\end{equation}
and the limit $\phi (-T)\rightarrow a_1,$ $\phi (T)\rightarrow a_2$ is
understood. Here $K(k),\Pi (\lambda k^2,k)$ are the complete elliptic
integral of the first kind and the third kind, respectively. To complete the
end-point integration, we choose the wave functions in the barrier as the
usual WKB form 
\begin{eqnarray}
\psi _{E_{cl},n}(\phi _i) &=&\frac{C\exp \left( -\int_{a_1}^{\phi _i}m(\phi )%
\stackrel{.}{\phi }d\phi \right) }{\sqrt{N(\tau _i)}}  \label{A2.11} \\
\psi _{E_{cl},n+1}(\phi _f) &=&\frac{C\exp \left( -\int_{\phi
_f}^{a_2}m(\phi )\stackrel{.}{\phi }d\phi \right) }{\sqrt{N(\tau _f)}}\equiv 
\frac{Ce^{-\Omega (\phi _f)}}{\sqrt{N(\tau _f)}}  \nonumber
\end{eqnarray}
and the amplitude is written 
\begin{eqnarray}
A_{f,i}^{E_{cl}} &=&\frac{C^2}{\sqrt{2\pi }}\int \frac{e^{-\int_{a_1}^{\phi
_i}m(\phi )\stackrel{.}{\phi }d\phi }e^{-\int_{\phi _f}^{a_2}m(\phi )%
\stackrel{.}{\phi }d\phi }}{N(\tau _i)}  \label{A2.12} \\
&&\times \left[ N^2(\tau _f)\int_{-T}^T\frac{d\tau }{N^2(\tau )m\left( \phi
_c(\tau )\right) }\right] ^{-\frac 12}e^{-S_c}d\phi _fd\phi _i  \nonumber
\end{eqnarray}
The end point integration $d\phi _i$ can be evaluated as 
\begin{equation}
\int \exp \left( -\int_{a_1}^{\phi _i}m(\phi )\stackrel{.}{\phi }d\phi
\right) \frac{d\phi _i}{\stackrel{.}{\phi }(\tau _i)}=2T  \label{A2.13}
\end{equation}
under the limit $\phi _i\rightarrow a_1.$ We expand $S_c$ and $\Omega
=\int_{\phi _f}^{a_2}m(\phi )\stackrel{.}{\phi }d\phi $ as power series of $%
\left[ \phi _f-\phi \left( T\right) \right] $ with the limit $\phi
_f\rightarrow a_2$ up to the second order for the Gaussian approximation,
i.e. 
\begin{eqnarray}
S_c &=&S_c\left( \phi (T),\phi (-T),T\right)  \label{A2.14} \\
&&+\frac 12\frac{\partial ^2S_c}{\partial \phi _f^2}|_{\phi _f=\phi \left(
T\right) }\left( \phi _f-\phi \left( T\right) \right) ^2+\cdots \cdots 
\nonumber
\end{eqnarray}
and 
\begin{equation}
\Omega =\frac 12\frac{\partial ^2\Omega }{\partial \phi _f^2}|_{\phi _f=\phi
\left( T\right) }\left( \phi _f-\phi \left( T\right) \right) ^2+\cdots
\cdots .  \label{A2.15}
\end{equation}
The transition amplitude then is 
\begin{eqnarray}
A_{f,i}^{E_{cl}} &=&\frac{2TC^2}{\sqrt{2\pi }}\int \left[ \frac 1{N^2(\tau
_f)\int_{-T}^T\frac{d\tau }{N^2(\tau )m\left( \phi _c(\tau )\right) }}%
\right] ^{\frac 12}  \label{A2.16} \\
&&\times e^{-W}e^{-2E_{cl}T}e^{-\frac 12\frac 1{N^2(\tau _f)\int_{-T}^T\frac{%
d\tau }{N^2(\tau )m\left( \phi \right) }}\left( \phi _f-\phi \left( T\right)
\right) ^2}d\phi _f  \nonumber \\
\ &=&2TC^2e^{-W}e^{-2E_{cl}T}  \nonumber
\end{eqnarray}
The renormalization constant $C$ is defined and evaluated as 
\begin{equation}
C=\left[ \frac{1/2}{\int_{a_1}^{a_2}\frac{d\phi }{\sqrt{\frac 2{m(\phi )}%
\left( E_{cl}-V\right) }}}\right] ^{\frac 12}=\left[ \frac \omega {%
4K(k^{\prime })}\right] ^{\frac 12}  \label{A2.17}
\end{equation}
where $k^{^{\prime }2}=1-k.$ Summing over contributions from one instanton
plus $n$ instanton-anti-instanton pair, the total amplitude is given by 
\begin{equation}
A_{f,i}^{E_{cl}}=\sum_{n=0}^\infty A_{f,i}^{(2n+1)}=e^{-2E_{cl}T}\sinh
\left( 2TCe^{-W}\right)  \label{A2.18}
\end{equation}
where the transition amplitude of one instanton plus $n$ pairs is seen to be 
\begin{equation}
A_{f,i}^{(2n+1)}=\int_{-T}^Td\tau _1\int_{-T}^{\tau _1}d\tau _2\cdots
\int_{-T}^{\tau _{2n}}d\tau [C]^{2n+1}e^{-(2n+1)W}e^{-2E_{cl}T}
\label{A2.19}
\end{equation}

The generalized level shift formula due to tunneling for
given energy $E_{cl}$ is obtained 
\begin{equation}
\Delta E_m=\left[ \frac \omega {4K(k^{\prime })}\right] ^{\frac 12}e^{-W}
\label{A2.20}
\end{equation}
and from which we have the energy level-spliting $2\triangle E_m\cos \pi
\Theta $ where $\Theta \pi =S[1-\frac{g\mu _BH}{2K_1S\sqrt{1-\lambda }}]$.

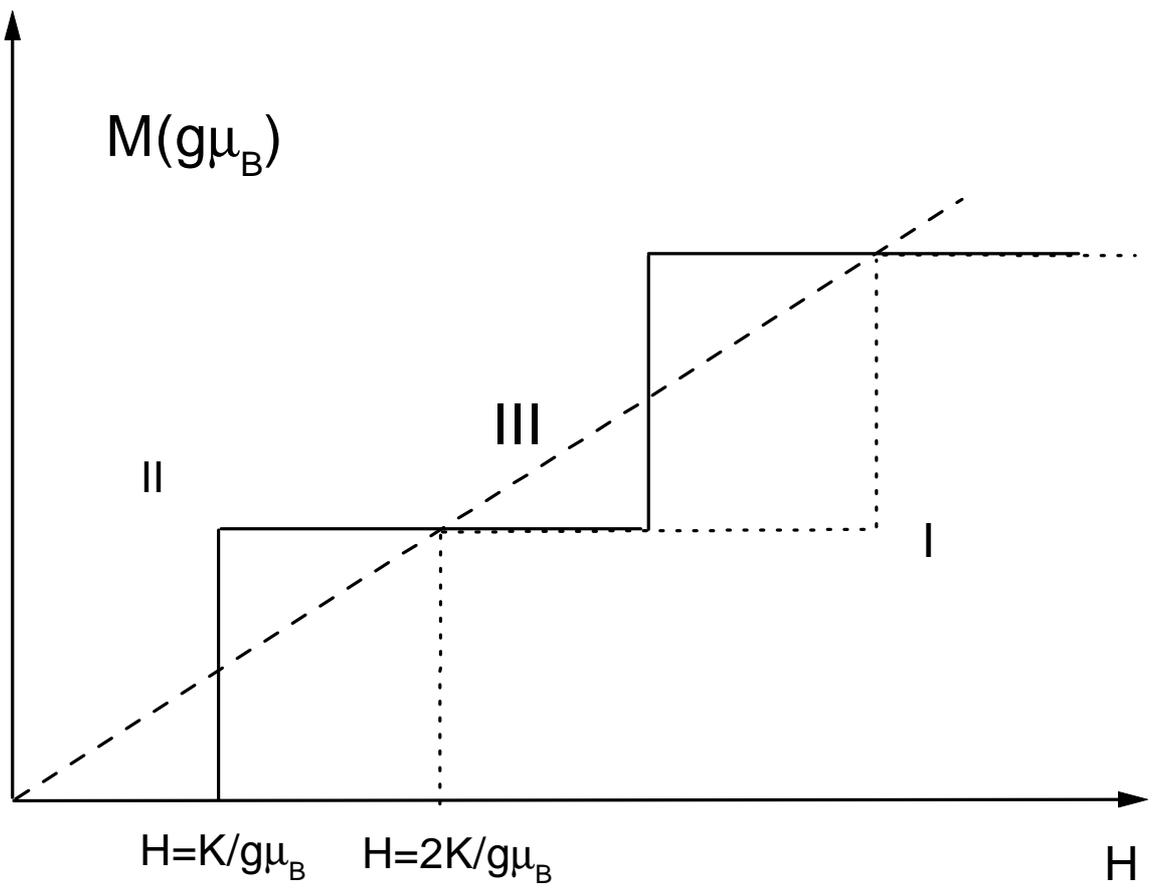
Fig. 1. Magnetization curves at zero temperature for a magnetic particle :
curve I (solid line) is magnetization curve for magnetic particles with
integer spin, curve II (dotted line) with half integer spin and curve III (
dashed line) is the classical one.

Fig.2 The energy level splitting $\triangle E_0$ of ground state for a
magnetic particle with $K_1=1$, $\lambda =0.2$ and (a) $S=16;$ (b) $S=16.5.$
The result of instanton method is plotted with the solid line and the
numerical result is shown by the dotted line.

Fig.3 The Schottky anomaly behavior of two-level system.

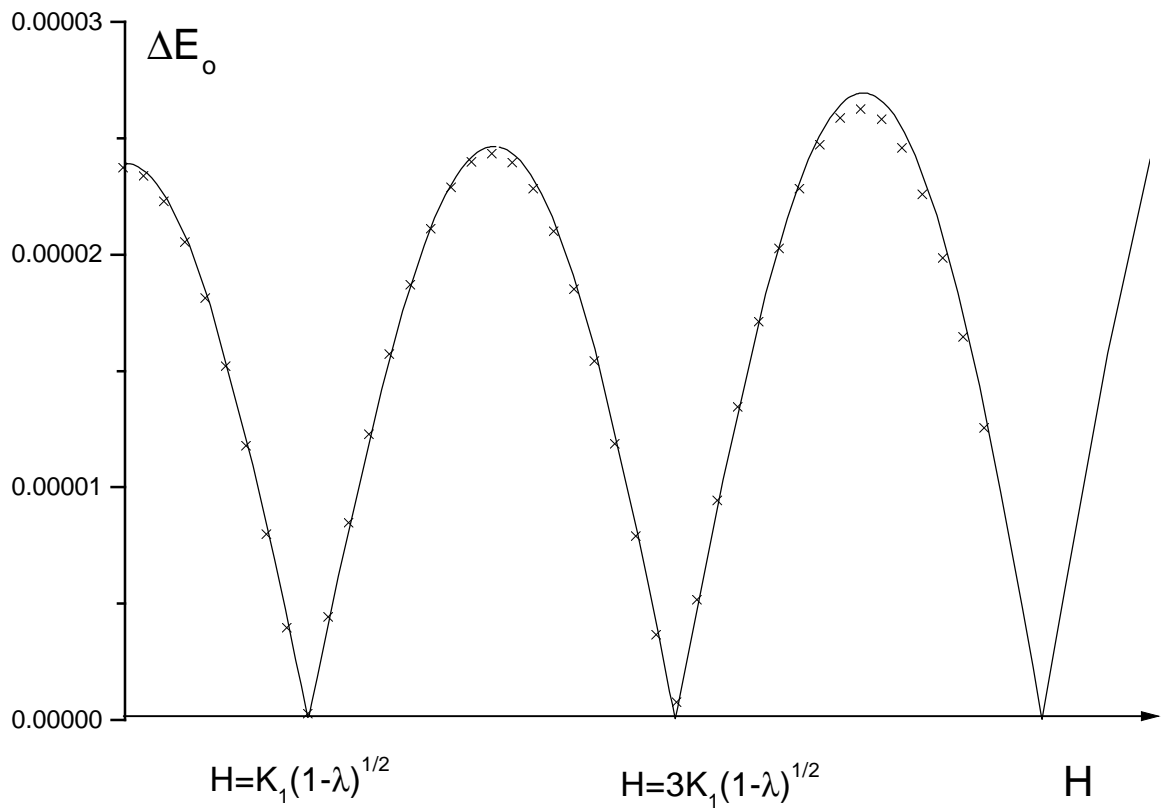
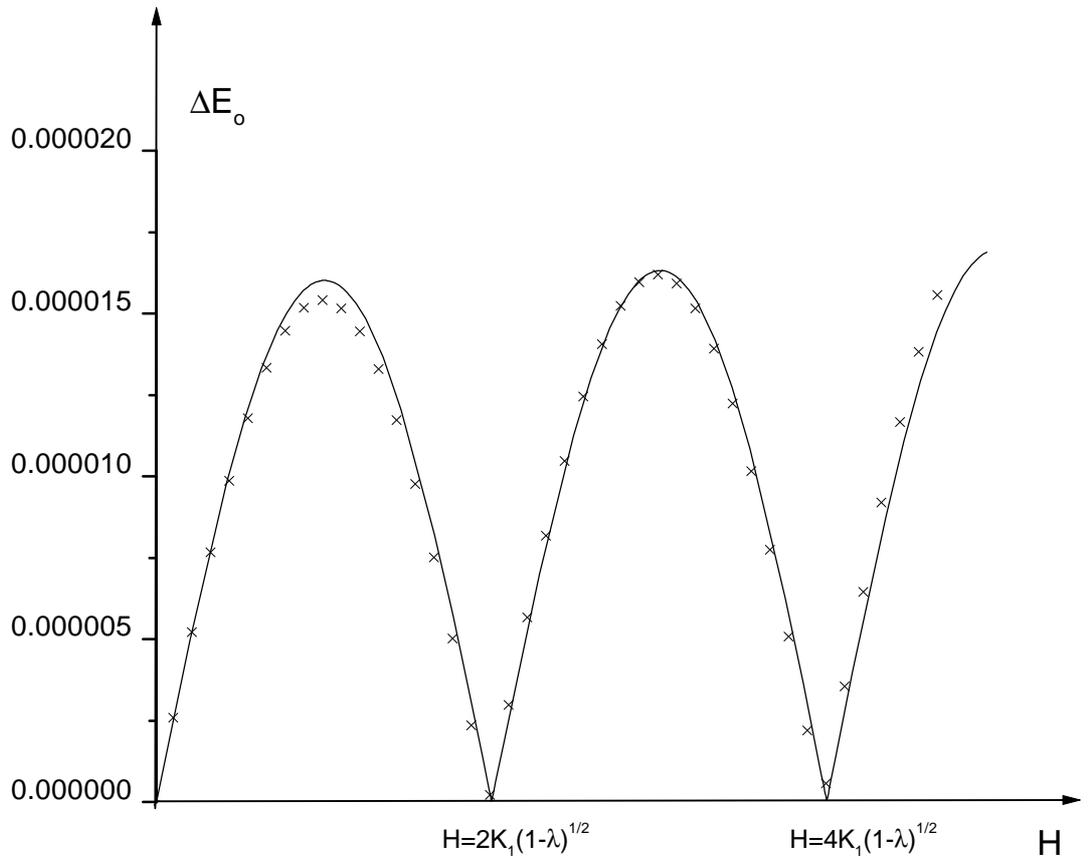
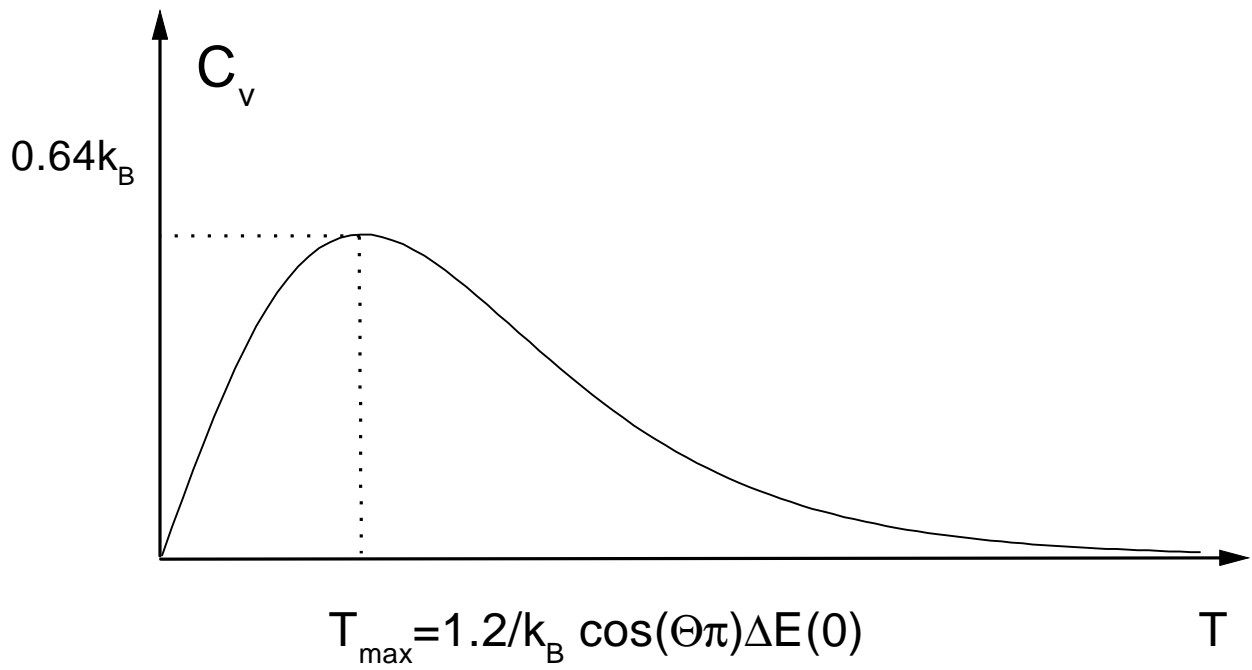
\end{document}